\documentclass[useAMS,usenatbib]{mn2e}
\usepackage{psfig,subfigure}

\title[A 3D Monte Carlo Photoionization Code for Modeling
Diffuse Ionized Gas]
 {A 3D Monte Carlo Photoionization Code for Modeling
Diffuse Ionized Gas}

\author[Wood, Mathis, \& Ercolano]
 {Kenneth Wood$^1$, John S. Mathis$^2$, and Barbara Ercolano$^3$ \\
 $^1$School of Physics \& Astronomy, University of St. Andrews,\\
North Haugh, St Andrews, KY16 9SS, Scotland;
kw25@st-andrews.ac.uk\\
 $^2$Astronomy Department, University of Wisconsin, \\
475 N. Charter Street, Madison, WI 53706; mathis@astro.wisc.edu\\
 $^3$Department of Physics \& Astronomy, University College London, \\
Gower Street, London, WC1E 6BT, UK,
be@star.ucl.ac.uk}

\date{Released 2002 Xxxxx XX}
\pagerange{\pageref{firstpage}--\pageref{lastpage}} \pubyear{2002}

\def\LaTeX{L\kern-.36em\raise.3ex\hbox{a}\kern-.15em
 T\kern-.1667em\lower.7ex\hbox{E}\kern-.125emX}

\begin{document}

\label{firstpage}

\maketitle

\begin{abstract}

We have developed a three dimensional Monte Carlo photoionization code
tailored for the study of Galactic H~{\sc ii} regions and the percolation of
ionizing photons in diffuse ionized gas. We describe the code, our
calculation of photoionization, heating \& cooling, and the
approximations we have employed for the low density H~{\sc ii} regions we
wish to study. Our code gives results in agreement with the Lexington
H~{\sc ii} region benchmarks. We show an example of a 2D shadowed region and
point out the very significant effect that diffuse radiation produced
by recombinations of helium has on the temperature within the shadow.

\end{abstract}


\section{Introduction}

A major characteristic of the interstellar medium (ISM) is that it is
clumpy, with filamentary structure over scales ranging from hundreds of
parsecs through sub-AU cloudlets. The spatial density of electrons is
approximately a power law over an astonishing 12 orders of magnitude
\citep{ars}, from $\sim$300 pc to $<$0.1 AU in the ISM. The magnetic
field of the Galaxy varies in a similar fashion on scales from 0.01 to
100 pc. Molecular clouds show a fractal (scale-free) spatial variation
of the intensity of CO (e.g., \citealt{fal}) and neutral H \citep{sl}.
Sub-parsec variations are shown by the variation of ionic column
densities along different ISM sightlines towards individual stars in
close binaries \citep{lmb}.  Large area velocity resolved surveys of the
molecular (e.g., \citealt{sjc}), neutral (e.g., \citealt{hb}), and
ionized \citep{r02} gas now allow us to probe the geometry, kinematics,
ionization, and temperature structure of the ISM.

Our development of Monte Carlo radiation transfer techniques to study
photon scattering, ionization, and radiative equilibrium dust
temperatures in complex geometries is motivated by these and the many
other observations that clearly require three dimensional modeling
analyses. In this paper we describe improvements and extensions to our
Monte Carlo photoionization code \citep{wl} that will enable us to
model the 3D temperature, ionization structure, emission line images,
and spectra of low density H~{\sc ii} regions and the Diffuse Ionized Gas 
(DIG), often called ``the warm ionized medium" or ``Reynolds layer''.
This diffuse ionized gas has a complicated structure and surely
requires three-dimensional modeling, both as regards the sources of
its ionization and the transfer of the ionizing radiation. How can
Lyman continuum photons from O and B stars percolate from their
midplane origins to ionize the high latitude gas? A non-uniform ISM is
required that provides low density paths for ionizing photons to reach
the halo.  Using a ``Str\"omgren volume'' ionization technique and a
statistical approach to clumping, \citet{mc} showed that O and B stars
can ionize the DIG (see also \citealt{ds94}; \citealt{dsf}). Using
Monte Carlo simulations for hydrogen ionization, \citet{c02} have
investigated ionization structures in a fractal ISM as proposed by
\citet{elm}. In Elmegreen's picture, the observed H$\alpha$ emission
may arise from the ionized surfaces of fractal clouds. This suggestion
clearly needs to be investigated with 3D radiation transfer
techniques.

Other features of the DIG that may require 3D analyses are the
temperature structure and relative ionization structure of He and H. The
temperature of the DIG, as determined from line intensity ratios (e.g.,
\citealt{rht}), appears to require additional heating over and above that 
provided by photoionization heating \citep{m00}. Among the possible heating
mechanisms, dissipation of energy in a turbulent medium has been
proposed \citep{ms}, definitely 3D! Another question is, can
clumping/shadowing explain the apparent lack of He$^0$ in H$^+$ regions
\citep{rt}? In this scenario, direct stellar photons are prevented from
ionizing He in shadow regions behind dense clumps, but diffuse ionizing
photons from H and He recombinations can percolate into the shadow
regions and ionize H, but not He.

Our work on the analysis of scattered light in fractal reflection
nebulae \citep{mww} highlighted the crucial role of geometry when trying
to determine dust properties from scattered light observations. Much of
what we know about abundances of elements comes from analysis of spectra
utilizing one-dimensional photoionization techniques (see e.g.
\citealt{s02}).  Over the last decade several three dimensional 
photoionization codes have been developed.  The code presented by 
\cite{bdg} uses the ``on the spot'' approximation for the 
diffuse radiation field, while that of \cite{gruen} employs an 
adaptive mesh grid and an iterative procedure to determine the 
contribution of the diffuse radiation field.  Monte Carlo techniques 
naturally include diffuse radiation within three dimensional systems and 
Monte Carlo photoionization codes have been described by \citet{erc} 
for the three-dimensional case and \citet{olr} for the one-dimensional case.

We have extended our basic hydrogen-only Monte Carlo photoionization code
\citep{wl} to treat the photoionization, heating, \& cooling for the
complex geometry expected in low density H~{\sc ii} regions and the diffuse
ISM. Our code is similar to those described by \citet{olr} and
\citet{erc}, but calculates the rates of photoionization and heating
``on the fly'', rather than calculating the mean intensity of the
radiation field (see also Lucy 1999). It also treats reprocessing of 
radiation in a different manner to that described by \citet{olr}, 
using ``photon packets'' rather than ``energy packets.''

In \S2 we describe our assumptions and approximations to model the DIG.
\S3 describes the Monte Carlo technique, our method of 
discretizing the radiation field into ``photon packets'' rather than 
``energy packets,'' and how we determine the diffuse radiation 
field using ratios of recombination rates.  
\S4, \S5, \& \S6 deal with our
photoionization and heating/cooling algorithms, \S7 presents the atomic
data used in our code, and \S8 compares our code with standard H~II
region benchmarks and gives an example of a 2D calculation.

\section{Approximations for the DIG}

We make the following assumptions and approximations for the opacity,
heating and cooling.

\begin{itemize}

\item We consider only radiation energetic enough to ionize H
($h\nu\ge$13.6 eV). Photons of lower energies are ignored until we
calculate the emergent intensities of certain spectral features after
the simulation is completed.

\item All ions are in the ground state when they absorb radiation or
recombine with an electron. This is the standard ``nebular
approximation.'' The low-lying levels giving rise to fine structure
lines might be an exception. Radiative transfer in these lines would
require consideration of the populations of the levels involved in
their production. We assume that the fine structure lines are
optically thin.

\item No H-ionizing photons are produced by recombinations to excited
levels of H or He. This is because $kT_e\ll \chi$(H$^0$) = 13.6 eV, the
ionization potential of H.

\item There is no collisional ionization or de-excitation. In effect,
this limits the densities we can study to be less than a few $\times
10^3$~cm$^{-3}$, which is generally above that encountered in the
Diffuse Ionized Gas and typical H~II regions, but within the typical
range of planetary nebulae.

\item There is no He$^{+2}$ because the stellar radiation
field has negligible luminosity above 54.4~eV, consistent with
observations of almost all H~{\sc ii} regions.

\item The opacity is due to continuous opacity from neutral H and He;
we ignore contributions from other ions because of their low
abundances. At present we ignore all other opacity sources (e.g.,
dust, UV metal line blanketing, etc.) except photoionization from
oxygen.

\item Photoionization and heating by the He~I Lyman~$\alpha$ line is
included in an ``on-the-spot'' approximation (see \S\ref{helya} for
a discussion of the physics).

\item We keep track of the abundances of the stages of ionization of
H, He, C, N, O, Ne, and S that have their ionization potentials
$\ge\,\chi({\rm H}^0)$ and $\le\,\chi(\rm{He}^+)$ = 54.4 eV. These are
H$^0$, He$^0$, C$^+$ -- C$^{+3}$, 
N$^0$ -- N$^{+3}$, O$^0$ -- O$^{+2}$, Ne$^0$ --
Ne$^{+3}$, and S$^+$ -- S$^{+3}$. These chemical elements are chosen
because some of their ions are important for the heating and cooling
rates.  It is straightforward
to include other elements, but they do not appreciably affect the H
and He ionization structure or contribute to the heating or
cooling. We assume that there is no S$^0$ because its ionization
potential is 10.36~eV, so it is ionized by the ambient stellar
radiation fields softer than $h\nu$ = 13.6 eV (see \citealt{ss} for
observations).

\item Heating arises solely from the photoionization of H and He.
Cooling is from recombinations of H and He, free-free emission, and
collisionally excited line radiation from N, O, Ne, S (see below).

\item We ignore dynamics, turbulence, and the effects of shocks and
ionization fronts.

\end{itemize}

It is straightforward to include more ions and the effects of dust
opacity. The assumption of no opacity from heavy elements may also be
easily relaxed.

\section{Monte Carlo Photoionization}

We want to determine the time-independent ionization and temperature
structure of a nebula with an assumed fixed density distribution. We do
this by Monte Carlo radiative transfer followed by iterations to find
the ionization structure that determines the opacity.

Within each iteration we consider packets of ionizing photons emitted
from the exciting star(s) of the nebula. We choose the frequency from
the probability distribution of the photon emission from the sources. We
follow each stellar photon packet through the density grid until it is
absorbed and causes photoionization. The subsequent recombination often
results in the production of another ionizing photon packet, thereby
forming the diffuse radiation field. The frequency of the new packet is
sampled from various atomic probability distributions. We first
determine if the packet was absorbed by H$^0$ or He$^0$, and then
determine the frequency of the re-emitted packet (see below). We
continue the Monte Carlo random walk of absorption/re-emission events
until the packet is re-emitted with $h\nu<13.6$~eV, at which time we
terminate it.

\subsection{Photon Packets versus Energy Packets}

Our approach differs somewhat from previous Monte Carlo simulations
(\citealt{olr}; \citealt{l99}; \citealt{bw}; \citealt{erc}) that
followed the {\em energy} of a packet, rather than its number of
photons. Considering energy is natural for investigating problems of
radiative equilibrium (e.g., scattering and thermal emission by dust, or
stellar atmospheres), since the energy that is absorbed is thermally
re-emitted with exactly the same energy but with a source function,
$S_\nu$, given by Kirchhoff's Law ($S_\nu\equiv j_\nu/k_\nu=B_\nu(T)$,
where $j_\nu$ is the emissivity and $k_\nu$ the absorption coefficient).
Kirchhoff's Law is applicable only when the populations of atomic levels
are in Local Thermodynamic Equilibrium and does not apply to nebulae.
The number of photons associated with an energy packet varies with its
frequency, so it is not as straightforward to apply atomic probabilities
in following the fate of an energy packet. The energy of an energy
packet is given by $\epsilon=L\,\Delta t/N$ and is kept constant for
radiative equilibrium codes, where $L$ is the stellar luminosity,
$\Delta t$ is the time covered by the simulation, and $N$ is the number
of packets used in the simulation. The energies of our packets are
$\epsilon=Q\,h\nu\,\Delta t/N$, where $Q$ is the ionizing photon
luminosity (units s$^{-1}$). The energy is not kept constant as the
packet changes frequency, but keeps its number of photons constant 
(see also Lucy 2003).

Since our initial goal is to determine the 3D ionization and temperature
structure, we do not follow the non-ionizing diffuse photons: they are
emitted with $h\nu<13.6$~eV and will escape from the nebula, so we
terminate the ionizing photon and start a new one from the star. We thus
save memory by not having to form probability distribution functions in
each cell for finding the frequency of non-ionizing photons. Rather, we
determine a converged ionization and temperature structure and then
employ a second code to integrate numerically through our resulting
emissivity grid to form the emergent spectrum and images in various
spectral lines.

Current desktop machines using 1Gb RAM can handle grids of around
$128^3$ cells for our code. The iterative photoionization and heating
calculation (described in \S4 and 5) updates the temperature and ionic
fractions of all species we are tracking in each cell throughout the
grid. At the end of each iteration we have a new ionization structure
and opacity grid for the next iteration.

\subsection{Emission Direction and Frequency}

In Monte Carlo techniques we must decide on the outcome of events that
depend on some physical variable, say $x$ (e.g., the frequency $\nu$),
with a given probability, $P(x)$. The important quantity is the
cumulative distribution function, $C(x)$, given by

\[
C(x)={{\int_{x_-}^{x}\,P(x^\prime)\,\rm dx^\prime}\over
\int_{x_-}^{x_+}\,P(x^\prime)\,\rm dx^\prime}\; ,
\]

\noindent where $x_-$ is the minimum value of $x$ and $x_+$ the maximum.
We find a randomly selected value of $x$ by casting a random
number\footnote{A new random number is generated by the subroutine
ran2(iseed) \citep{pr} each time $\xi$ appears in this paper. A new
integer, iseed, is automatically generated by the subroutine after
each entry. }, $\xi$, in the range [0,1] and inverting the relation
$\xi=C(x)$.

We assume the density structure of our nebular simulation, possibly
quite complex as appropriate to the ISM. We begin the Monte Carlo
radiation transfer by emitting photon packets from the exciting star(s)
with an initial guess for the ionization structure. The initial guess
for the ionization structure does not affect the final converged
solution, but it does affect the number of iterations required for
convergence. For isotropic emission, the random directions are

\begin{eqnarray}
&\cos\theta=2\xi-1\\
&\phi=2\pi\xi\; ,
\end{eqnarray}

\noindent where $\theta$ and $\phi$ are the usual spherical polar
coordinates. The direction cosines for the initial photon trajectories
are then given by

\begin{eqnarray}
&n_x=\sin\theta\cos\phi\cr
&n_y=\sin\theta\sin\phi\cr
&n_z=\cos\phi\; .\label{cph}
\end{eqnarray}

The photon frequency must be sampled to reproduce the ionizing spectrum
of the source, $L(\nu)$, which may be a blackbody or model atmosphere.
We pre-tabulate the cumulative photon distribution function,

\begin{equation}
C(\nu_i)={ {\int_{\nu_{H^0}}^{\nu_i} L(\nu)/h\nu\,{\rm d}\nu} \over
 {\int_{\nu_{H^0}}^{\infty} L(\nu) /h\nu\,{\rm d}\nu} }\; .
\end{equation}

\noindent A random frequency is chosen by solving for $\nu$ in $\xi=C(\nu)$ 
by interpolating in the pre-tabulated cumulative distribution function
$C(\nu_i)$. From $\nu$ we calculate the photoionization cross sections
$a_\nu({\rm H}^0)$ and $a_\nu({\rm He}^0)$ from analytic formulae. The
optical depth, $\tau_\nu$, along a path length $S$ is

\begin{equation}
\tau_\nu = \int_0^S[n({\rm H}^0)a_{\nu}({\rm H}^0) +
n({\rm He}^0)a_{\nu}({\rm He}^0)]\,{\rm d}s\; .
\end{equation}

\noindent where $n({\rm H}^0)$ and $n({\rm He}^0)$ vary from cell to
cell. We generate a random optical depth by $\tau =-\log\xi$ and
integrate through the opacity grid until the distance $S$ has been
reached. If the distance $S$ lies outside the simulation grid the photon
has exited our simulation and we start another packet from the star.

\subsection{The Diffuse Radiation Field}

On finding the position associated with the randomly chosen optical
depth, we decide if the photon packet is absorbed by H$^0$ or He$^0$.
The packet is absorbed by H$^0$ if $\xi\le n({\rm H}^0)a_\nu({\rm
H}^0)/[ n({\rm H}^0)a_\nu({\rm H}^0)+ n({\rm He}^0)a_\nu({\rm He}^0)]$,
and by He$^0$ otherwise. If absorbed by H$^0$, it will be re-emitted
either as a Lyman continuum photon (and subsequently tracked with the
Monte Carlo technique) or terminated. If absorbed by He$^0$ there are
several energy channels into which it can be re-emitted as described
below. Once a photon packet has been absorbed we re-emit it with a
frequency chosen using the following algorithm.

\subsubsection {H~I Emission}\label{hem}

The probability that the photon is emitted in the hydrogen Lyman
continuum is

\begin{equation}
P({\rm H~I~Ly~c})={ {\alpha_1({\rm H}^0, T_e)} \over {\alpha_A({\rm
H}^0\;
,T_e)} }\; ,\label{phlyc}
\end{equation}

\noindent where $\alpha_1$ is the recombination coefficient to the
ground level and $\alpha_A$ the coefficient to all levels, including $n$
= 1. The packet emerges in the Lyman continuum if $\xi\le P({\rm
Ly~c})$. If the packet is chosen to be re-emitted in the Lyman
continuum, we sample its frequency using the $C(\nu)$ derived from the
Saha-Milne emissivity:

\begin{eqnarray}
&{C}_{\rm{Lyc}}(\nu, T_e)=\int_{\nu_0}^\nu (j_\nu/h\nu)\,
{\rm d}\nu/\int_{\nu_0}^\infty (j_\nu/h\nu)\,{\rm d}\nu\; ;\\
&j_{\nu}={{2h\nu^3}\over{c^2}}\,
\left({h^2}\over{2\pi\, m_e\, k\, T_e}\right)^{3/2}\cr
&\times\; a_{\nu}({\rm H^0})
{\rm e}^{-h(\nu -\nu_0)/kT_e} n({\rm H}^+)n_{\rm e} \; .\label{chlya}
\end{eqnarray}

We pre-tabulate $C_{\rm{Lyc}}(\nu,\,T_e)$ and from $\xi = C(\nu, T_e)$ we
determine $\nu$ from $\xi$ by interpolating in the table. The packet is
now re-emitted with a direction chosen from an isotropic distribution
(Eqs.~1 and 2).

At the start of each iteration we form a 3D grid that contains $P({\rm
H~I~Ly~c})$ and use this grid to decide the fate of photon packets
absorbed by H$^0$. Similar grids are formed for the He emission
probabilities as described in the next section. These probability grids
are updated after each iteration since the recombination rates are
temperature dependent.

\subsubsection {He~I Emission}\label{HeI}

An absorption by He is followed by recombination and possibly by
radiative cascades into either the metastable levels $2^1S$ or $2^3S$,
to $2^1P$ (the upper level of Lyman $\alpha$), or to the ground level
$1^1S$. The probabilities for emission in these four channels are as
follows:

\begin{enumerate}

\item He~I Lyman continuum ($h\nu> 24.6$~eV) is analogous to that
given for H in \S\ref{hem}.

\item
The probability of emission as a 19.8~eV photon packet from the
transition $2^3S\rightarrow 1^1S$ is

\begin{equation}
P(19.8\,{\rm eV})={ {\alpha_{2^3S}^{\rm eff}({\rm He}^0, T_e)}
\over {\alpha_A({\rm He}^0, T_e)} }\; .
\end{equation}

\noindent The {\it effective} recombination rate takes into account all
means of populating the given level, both direct recombination and via
cascades from higher levels, from which low energy photons are emitted.

\item The probability of emission from the He~I two-photon continuum from
$2^1S$ is

\begin{equation}
P({\rm He~I}\,2q)={ {\alpha_{2^1S}^{\rm eff}({\rm He}^0, T_e)}
\over {\alpha_A({\rm He}^0, T_e)} }\; .
\end{equation}

Note that $\alpha_{2^1S}^{\rm eff}({\rm He}^0, T_e)$ does not include
de-excitations from $2^1P$, which will be discussed in \S4.1

A fraction 0.28 of the photons in this emission channel is able to
ionize H$^0$, so the number of photons emitted per decay is 2(0.28) =
0.56 photons. After determining that the packet is to be emitted in the
2-photon continuum, we see if $\xi \le 0.56$. If so, we choose a
frequency from a pre-tabulated $C(\nu)$ for two-photon emission
(\citealt{dvd}, Table~II).

\item The probability of emission as a He~I Lyman $\alpha$ photon
($h\nu=21.2$~eV) is

\begin{equation}
P({\rm He\,I\,Ly\,\alpha})={ {\alpha_{2^1P}^{\rm eff}({\rm He}^0,
T_e)}\over {\alpha_A({\rm He}^0, T_e)} }\; .\label{hep4}
\end{equation}

\citet{olr} and \citet{erc} followed energy packets of several of the He~I
Lyman lines, assuming that the various He Lyman lines are absorbed only
by the H$^0$ continuous opacity, propagating through the grid and
maintaining their identity without degrading into Lyman~$\alpha$. We do
not follow this prescription because resonance lines will be scattered
many times without diffusing far in space. Following each scattering
from $n^1P$ ($n\ge2$), there is a finite probability of re-emission into
a lower excited level rather than to 1$^1S$. Subsequent emissions from
that excited level cascade downwards and populate either $2^1S$ or
$2^1P$. Conversion to $2^1S$ is followed by He~I 2-photon continuum
emission. We discuss our treatment of the $2^1P$ in \S\ref{helya}.

\end{enumerate}

To form the various probabilities, we use analytic fits to the
temperature dependent recombination rates for H and He. These rates,
along with all other atomic data, are presented in \S\ref{atdata}. Since
there are only four He~I levels with appreciable populations, we have

\begin{equation}
\alpha_A = \alpha_1+\alpha_{2^1S}^{\rm eff}+\alpha_{2^1P}^{\rm eff}+
\alpha_{2^3S}^{\rm eff} \; .
\end{equation}

\noindent We also assume that recombinations to triplets produce only the 
19.8~eV 2$^3S\rightarrow 1^1S$ transition. We do not follow $2^3P\rightarrow
1^1S$, as did \citet{olr}, since at low densities 2$^3P$ converts to
2$^3S$ via 10830{\AA} emission.

\section{Photoionization}

We must calculate the photoionization, heating, and cooling in order to
update the ionic fractions and electron temperature after each
iteration.

The basic photoionization equations, in the notation of \citet{o89}, are

\begin{equation}
n(X^{+i})\int_{\nu_i}^\infty {{4\pi J_{\nu}} \over {h\nu}}
a_{\nu}(X^{+i}){\rm d}\nu = n(X^{+i+1})n_{\rm e}\alpha(X^{+i},T_e),
\label{ioneqn}
\end{equation}

\noindent where $n(X^{+i})$ and $n(X^{+i+1})$ are the number densities
(per cm$^3$) of successive stages of ionization, $\nu_i$ is the
threshold for photoionization from $X^{+i}$, $n_{\rm e}$ is the electron
number density, and $\alpha(X^{+i},T_e)$ is the recombination
coefficient to all levels of $X^{+i}$. In our code the photon luminosity
of the sources, $Q$, is divided into $N$ packets, so we calculate the
integral on the left of eqn. (\ref{ioneqn}) by

\begin{equation}
I^P_{X^{+i}}=\int_{\nu_i}^\infty {{4\pi J_{\nu}} \over {h\nu}}
 a_{\nu}(X^{+i}){\rm d}\nu = { {Q} \over {N \,\Delta V} }
\sum l a_{\nu} (X^{+i})\; .
\end{equation}

\noindent where the summation is over all path lengths through the
volume $\Delta V$ for photons in the frequency range ($\nu, \nu + {\rm
d}\nu$), and $l$ is the path length of each photon counted in the sum
(Lucy 1999, 2003).

Thus, we do not explicitly calculate the mean intensity $J_{\nu}$ 
throughout the grid in order to compute the integral of 
$4\pi J_{\nu}a_{\nu}(X^{+i})/h\nu$ for each ion considered. To save 
$J_\nu$, we would need storage for each grid cell with as many frequencies 
as being considered.

We add the contribution of each ionizing photon packet, regardless of
stellar or nebular origin, to the sum of $l\, a_{\nu}(X^{+i})$
maintained for each ion, within each cell in the grid. At the end of the
iteration we have a 3D array for each ion that contains the integral
$I^P_{X^{+i}}(x, y, z)$. The photoionization rate is then $n(X^{+i})\,
I^P_{X^{+i}}$ within each cell. So long as the number of ions being
tracked is less than the typical number of frequencies required for an
accurate representation of the mean intensity, our technique will save
significantly on memory storage. We use the above equations to solve the
ionization balance of most ions that we follow. Exceptions to this are
our treatment of He~I Ly$\alpha$ photons that can ionize H$^0$, and also
charge exchange that couples the ionization of some ions to H and He.

We now outline our treatments of the ionization balance for these
situations.

\subsection{He~I Lyman $\alpha$}\label{helya}

We do not need to follow the diffusion of the He~I Ly$\alpha$ resonance
line ($2^1P\rightarrow 1^1S$) in space and frequency. The level $2^1P$
is depopulated either by emission of a He~I Ly$\alpha$ photon
($2^1P\rightarrow 1^1S$) or a 2.06$\mu$m photon ($2^1P\rightarrow
2^1S$). Due to the large resonant line opacity, the He~I Ly$\alpha$ will
be scattered locally many times without traveling far. It will
eventually be absorbed locally (``on the spot'') by H$^0$ or else decay
to $2^1S$.

Note that the ``on the spot'' approximation is used due to the large
He~I Ly$\alpha$ resonant opacity and not the H$^0$ opacity.

What is $P({\rm H_{OTS}})$, the probability of the He~I Ly$\alpha$
photon being absorbed by H$^0$ as opposed to the transition
$2^1P\rightarrow 2^1S$ occurring?  The average number of resonant
scatterings before decay to 2$^1S$ is $A_{2^1P,1^1S}/A_{2^1P,2^1S}=
914$ (\citealt{bss}). The mean free path between scatterings, $l_{\rm
scat}$, is $[n({\rm He}^0)\,a_\nu({\rm Ly}\alpha)]^{-1}$, where the
line opacity is \\

\begin{equation}
a_\nu({\rm Ly}\alpha)=(\pi^{\frac{1}{2}}e^2/m_ec)\ f_{21}\,
\exp[-(\Delta\nu/\Delta\nu_D)^2]/\Delta\nu_D
\ , \label{lyal}
\end{equation}

\noindent where $f_{21}$, the oscillator strength of the
transition, is 0.29. $\Delta\nu/\Delta\nu_D$ is the displacement of the
frequency of the photon from the line center, in Doppler widths. This
displacement is determined by the speed the atom had before
recombination, and so has a Maxwellian probability distribution (if we
neglect any non-thermal components in the {\em local} velocity
distribution). The Maxwell-Boltzmann distribution can be written $f_{\rm
MB}(v)\propto v^2\,\exp[-(v/v_{\rm th})^2]$ and $\Delta\nu/\Delta\nu_D
=(v/v_{\rm th}$), with $v_{\rm th} = (2kT/M)^{1/2}$. We average over the
line profile:

\begin{eqnarray}
&\langle\exp[-(\Delta\nu/\Delta\nu_D)^2]\rangle\\
&={{\int_0^\infty \exp[-(v/v_{\rm th})^2]\,
v^2 \exp[-(v/v_{\rm th})^2]\,{\rm d}v}\over
{\int_0^\infty v^2\,\exp[-(v/v_{\rm th})^2]\,{\rm d}v}} \nonumber \\
&= 2^{-3/2}\ .
\end{eqnarray}

The number of absorptions by H$^0$ during the 914 scatterings is 914
$l_{\rm scat}\,n({\rm H}^0)\,a_\nu({\rm H}^0)$, with $a_\nu({\rm H}^0)=
1.9\times 10^{-18}$ cm$^2$. The probability of the photon being absorbed
by H$^0$ is

\begin{eqnarray}
P({\rm H_{OTS}})={{n({\rm H}^0)a_{\nu}
({\rm H}^0)}\over{
[n({\rm H}^0)\,a_{\nu}({\rm H}^0)+n({\rm He}^0)a_{\nu}({\rm
He}^0)/914]}}
\nonumber \\
= \left(1+0.77 f({\rm He}^0)/[f(\rm H^0) (T/10^4\,\rm K)^{1/2}]
\right)^{-1}\ ,
\end{eqnarray}

\noindent where $f({\rm He}^0)$ and $f({\rm H}^0)$ are the neutral
fractions of He and H, respectively, and He/H = 0.1 in abundance has
been assumed. The $f({\rm He}^0)/f({\rm H}^0)$ favors He, especially for
softer exciting stars. For instance, the inner parts of the nebula
surrounding a 40000~K black body favor He$^0$ by a factor of $\sim$7, so
80\% of transitions to 2$^1P$ are converted to 2$^1S$ rather than being
absorbed by H$^0$. For harder exciting spectra, the $f({\rm
He}^0)/f({\rm H}^0)$ is $<$2, and a significant number of Ly$\alpha$
photons are absorbed by H$^0$. Heating by Ly$\alpha$ is significantly
greater than from 2-photon emission. However, the (2$^3S \rightarrow
1^1S$) photons are far more important than any from the singlets as
regards both ionization and heating.

In the ``on-the-spot'' treatment, local absorptions of Ly$\alpha$ are
exactly balanced by local emissions. The rate of ``on-the-spot''
absorptions by H$^0$ is $P({{\rm H}_{\rm OTS}})$ times the rate of
recombinations into He$^0(2^1P$), or $P({{\rm H}_{\rm OTS}}) n({\rm
He}^+)\, n_{\rm e}\alpha^{\rm eff}({\rm He}^0, 2^1P)$. These absorptions
by H$^0$ are also given by $n({\rm H}^0)\,a_{\nu}({\rm
H}^0)[4\pi\,J_\nu({\rm Ly}\alpha)/h\nu]$. By equating these rates, we
obtain a modified ionization balance equation for H:

\begin{eqnarray}
&n({\rm H}^0)\int_{\nu_i}^\infty (4\pi J_\nu/h\nu)
 a_{\nu}({\rm H}^0){\rm d}\nu =\cr
&n_{\rm e}[n({\rm H}^+)\alpha({\rm H}^0,T_e) -
P({{\rm H}_{\rm OTS}})n({\rm He}^+)\alpha_{2^1P}^{\rm eff}]\; ,
\end{eqnarray}

\noindent where the $J_\nu$ does not include He~I Ly$\alpha$, which is
contained in the final term on the right side of the equation.

\subsection{Charge Exchange}

The rate of exchange of an electron between an ion ($X^{+i}$) and H$^0$
or He$^0$ can be comparable to or exceed the radiative recombination
rates. With our adopted charge exchange cross sections \citep{kf}, the
rate of transitions into N$^{+2}$, O$^0$, and O$^+$ are seriously
affected or dominated by exchange with H$^0$. The opposite effect, in
which charge exchange dominates the ionization of the lower stage, takes
place for O$^0$. A result is that in regions where O$^0$ is important
(i.e., H$^0$ is appreciable), (O$^0$/O$^+) \sim$ 9/8 (H$^0$/H$^+$)
for all $T_e$ because the ionization potential of O$^0$ is almost the
same as for H$^0$ (see \citealt{o89}, p. 42 for a discussion.) For
N$^0$, charge exchange is not dominant but not negligible. We included
the reactions in the rates for reactions resulting in N$^+$ and O$^+$ as
well.

With charge exchange included, the (O$^0$/O$^+$) equation becomes

\begin{eqnarray}
&n({\rm O}^0)\,[n_e\,\int 4\pi\,J_\nu/h\nu\, {\rm d}\nu + n({\rm H}^+)\,
C_{\rm exch}({\rm O}^0,\,{\rm H}^+)] = \nonumber\\
&n({\rm O}^+)\,[n_e\,\alpha_{\rm rec}({\rm O}^+)
+ n({\rm H}^0)\,C_{\rm exch}({\rm O}^+,\,{\rm H}^0)]\; .\nonumber
\end{eqnarray}

The situation regarding charge exchange with He$^0$ is less clear
because the cross section is very difficult to compute. We account for
charge exchange with He$^0$ for interactions resulting in N$^+$,
N$^{+2}$, and O$^+$.

\section{Heating \& Cooling}

The temperature is determined by finding the root of $G(T)= {\cal
L}(T)$, where $G$ and ${\cal L}$ are the heating and cooling rates due
to the processes described below (see Chapter~3
of \citealt{o89}).

\subsection{Heating}

We consider only heating by photoionization of H and He and 
calculate the heating rate in each grid cell in the same manner as we
calculated the photoionization rate. For example, for photoionization
of He the heating rate is

\begin{equation}
G({\rm He}^0) = \int_{\nu_{{\rm He}^0}}^\infty {{4\pi J_{\nu}} \over
{h\nu}}
 a_{\nu}({\rm He}^0)\,
 h(\nu - \nu_{{\rm He}^0})\,{\rm d}\nu\; ,
\end{equation}

\noindent so we form integrals analogous to the photoionization rate

\begin{eqnarray}
&I^H_{\rm He}=\int_{\nu_{{\rm He}^0}}^\infty {{4\pi J_{\nu}} \over
{h\nu}}a_{\nu}({\rm He}^0)\,
 h(\nu - \nu_{{\rm He}^0})\,{\rm d}\nu=\cr
&[Q/ (N \,\Delta V)]
\sum l a_{\nu} ({\rm He}^0)
(h\nu - h\nu_{{\rm He}^0})\; .
\end{eqnarray}

As with photoionization, we assume ``on-the-spot'' heating by He~I
Ly$\alpha$ packets, so a term $P({\rm H_{OTS}})n({\rm He}^+)n_e
\alpha^{\rm eff}_{2^1P}({\rm Ly}\alpha)
[h\nu({\rm Ly}\alpha)-\chi(\rm H^0)]$ is added to the heating per
volume of each cell.

\subsection{Cooling from Recombination and Free-Free Radiation}

Cooling rates as a function of temperature, $\cal L$, due to recombination 
of H$^+$ and He$^+$ are taken from Hummer (1994) and Hummer \& Storey 
(1998).  
The cooling rate due to free-free radiation is given by \citet{o89}:

\begin{equation}
{\cal L}_{\rm ff}=1.42\times 10^{-27}\,[n({\rm H}^+)+n({\rm He}^+)]\,
n_e\,g_{\rm ff}\sqrt T_e \; ,
\end{equation}

\noindent where we use the \citet{kwh} fit to
the free-free Gaunt factor data in \citet{spitzer}

\begin{equation}
 g_{\rm ff}=1.1+0.34\,\exp{[-(5.5-\log T_e)^2/3]} \; .
\end{equation}

\subsection{Other Cooling}

We include collisional excitation of the lowest five levels of the
ions of N$^0$, N$^+$,O$^0$,O$^+$,O$^{+2}$, Ne$^{+2}$, S$^+$, and
S$^{+2}$, and of 2 levels of N$^{+2}$ and Ne$^+$, as in other
photoionization codes (e.g., \citealt{erc}).
For this cooling by collisionally excited forbidden lines we use the
Einstein A values and collision strengths
from the compilation by \citet{pp95} as
updated on A. Pradhan's webpage.\footnote{
http://www-astronomy.mps.ohio-state.edu/$\sim$pradhan/}

\section{Some Code Details}

After each iteration, we simultaneously solve for the ionization and
temperature using the equations outlined above. 
The code will converge quickly if we have a good
first guess of the ionization and temperature structure. However, for
the 3D problems that we wish to address, such a guess will be very
difficult to make. We therefore start the first iteration assuming that
H and He are fully ionized, so all stellar photons exit the grid on the
first iteration, but do contribute to the mean intensity counters. If we
started with a neutral grid, photon packets could not diffuse far from
their point of origin due to the large neutral opacity and convergence
would be very slow, if at all.

As in \citet{erc}, we run several iterations with a small number of
photon packets to get a rough idea of the ionization and temperature
structure: $10^6$ packets for the first eight iterations, up to $10^8$
for higher iterations $\ge$20. Our models are well converged within 12
iterations; the higher iterations with large numbers of photon packets
provide higher signal-to-noise for the ionization and temperature
structure.

In 3D simulations, the diffuse field ionizes walls strongly inclined
to the direction to the star. We find a good early approximation to
the diffuse field by initially setting all cells to be fully ionized,
thereby predicting the diffuse radiation field if the nebula were
fully ionized. For the next iteration, if the stellar radiation cannot
maintain a high degree of ionization the outer material becomes almost
neutral, since the diffuse field is much weaker than the stellar
field.  The structure reaches a good approximation to the eventual
diffuse field after about five iterations. If the shadowed regions
were allowed to turn neutral before the diffuse field had converged,
we would face the same convergence problem (with each new iteration
ionizing a small zone of previously neutral material) described above.

We run our code on a single desktop machine within a reasonable CPU
time (see \citealt{erc} for a discussion of code speed up via
parallelization), running $\sim$1.0M photon packets per minute on a 2.4
GHz processor.

\section{Atomic Data}\label{atdata}

We use standard references for the line and continuum emissivities
(\citealt{b72}; \citealt{bm70}; \citealt{sh};
\citealt{bss}). We take the photoionization cross section for H from
the fitting formula of \citealt{vea96}. The recombination rate to
all levels $\alpha_A({\rm H}^0,T_e)$ is given by the fitting formula
in \citet{vf96}. For the recombinations to the ground
state, $1^2S$, we fitted the temperature dependent rates in
Osterbrock (1989, Table 2.1), with

\begin{equation}
\alpha_1({\rm H}^0,T_e)=1.58\times 10^{-13}(T_e/10^4 K)^{-0.53}\; .
\end{equation}

For He we use the \citet{vea96} formula for the photoionization cross
section. For the recombination rates we use the fitting formulae in
Table~1 of \citet{bss} that give us the direct recombination rates to
the ground level, $1^1S$, and the effective rate to $2^3S$. We
calculated the effective rates to $2^1P$ and to $2^1S$, ignoring the
$2^1P\rightarrow 2^1S$ transition because we separately consider its 
competition with absorption by H (see the discussion in \S4.1). 
The rates to $2^1P$ and $2^1S$
were formed by adding the direct recombination rates (\citet{bss},
Table~1), effective rates for $n>5$ (Table~3), and effective rates
from all lines for $3\le n\le 5$ that connect to the appropriate lower
level (Table~5). In summary,

\begin{eqnarray}
\alpha_1({\rm He}^0,T_e)=1.54\times 10^{-13}(T_e/10^4)^{-0.486}\cr
\alpha_{2^1S}^{\rm eff}({\rm He}^0, T_e)=
2.06\times 10^{-14}(T_e/10^4)^{-0.676}\cr
\alpha_{2^1P}^{\rm eff}({\rm He}^0, T_e)=
4.17\times 10^{-14}(T_e/10^4)^{-0.861}\cr
\alpha_{2^3S}^{\rm eff}({\rm He}^0, T_e)=
2.10\times 10^{-13}(T_e/10^4)^{-0.778}\; ,
\end{eqnarray}

\noindent thus enabling us to form the probabilities for reprocessing
packets by He (\S\ref{HeI}).

The ionization cross section for the ions we use are taken from
\citet{vea96} and the radiative recombination rates are from
\citet{vf96}. Also important in the recombination are dielectronic
recombination rates, for which we use values determined by Nussbaumer
\& Storey (1983, 1984, 1987) for N, O, and Ne. The dielectronic
recombination rates for sulphur have not been calculated, and we
normally use those recommended by \citet{a91} for the first four
ionization stages of S: $3\times 10^{-13}$, $3\times 10^{-12}$,
$1.5\times 10^{-11}$, and $2.5\times 10^{-1}$, which are the mean rates
for the first four atoms/ions of C, N, and O as discussed by
\citet{a91}. The charge exchange rates are taken from \citet{kf} for
reactions with H$^0$ and from \citet{ar} for those with He$^0$.

\section{Some Results and Benchmarks}

\subsection{Lexington H~II Region Benchmarks}

Any 3D photoionization code must reproduce two benchmark tests of
blackbody exciting stars ionizing uniform spherical nebulae of a
specified composition. The benchmarks have been discussed extensively
in workshops (\citealt{f95}, \citealt{p01}; see these for the details
of the input parameters). Results of many codes are given in Table 1
(for a 40kK central star) and Table 2 ($T_*$ = 20kK). The authors of
the other codes are given in the footnote to Table 1. The dielectronic
recombination coefficients assumed for S are zero for S$^{+2}$ and
those in \citet{a91} for S$^{+3}$ and S$^{+4}$. We believe these were
adopted by most of the models in each table. Some models in the
benchmark took the optical depths of fine structure lines into
account, but we did not. Our results are given in the last column in
each table.

Figure~1 shows plots of ionization and temperature with radius for the 
$T_\star$ =40000~K benchmark and compares our results (crosses) with 
the Monte Carlo photoionization code {\sc mocassin} \citep{erc} (squares) and 
the widely used code {\sc cloudy} \citep{f94} (lines).  Figure~ 2 presents the 
same quantities for the $T_\star$ = 20000~K benchmark.  In comparing with 
{\sc mocassin}, we set up the density grids 
to have the same spatial resolution: 
our code used a grid of $65^3$ cells and {\sc mocassin}, which utilizes 
the symmetry of the benchmark, performed the radiation transfer in an 
$33^3$ octant.  For these simulations our code used $10^8$ photon packets 
and {\sc mocassin} used $6.4\times 10^7$ energy packets in the final (20th) 
iteration, though both codes had essentially converged by iteration 14.  
Among the three codes, the ionization fractions are all in very good 
agreement.  Because of our 65$^3$ grid resolution, our code
produces slightly more ionization at the outer edge.  There is a larger 
spread in the predicted temperatures for the hot benchmark, with our code 
being slightly hotter in the inner portions of the nebula and {\sc mocassin} 
being slightly cooler towards the edge of the nebula.  
The agreement for the cool benchmark temperatures is very good.

The large spread of some ionization fractions (e.g., O$^{+2}$/O in 
Fig.~2) illustrates a property of Monte Carlo codes unless special steps 
are taken.  The ionization potential of O$^+$ is 35.1 eV;
an ionizing photon is so rare that even with $10^8$ photon packets there
is considerable statistical noise. If an accurate prediction of the
(O$^{+2}$/O) were one of our goals, we could have forced energetic
stellar photons to be emitted and then correct the results by the
known probabilities of the events we forced.

The cubic nature and limited spatial resolution of our grid makes the
comparison problematic for one ion, O$^0$, mainly found in the very
outermost regions of the nebula. In these regions, (O$^0$/O$^+$) = 9/8
H$^0$/H$^+$ because of a resonant charge exchange between O$^+$ and
H$^0$. There is a rather sharp peak in the temperature because of the
hardening of stellar radiation caused by selective absorption within
the nebula, and a steep temperature gradient because of the large
opacity.  With better spatial resolution, the volume occupied by one
cell would contain regions of almost neutral hydrogen, so the
temperature assigned to the whole cell, and the strength of [O~I]
$\lambda$6300, are of limited meaning. We present results from our
cells with H$^0$/H $\le$ 0.25; the cells more neutral than this amount
occur in the outer $<$1\% of the radius ($r_{\rm outer} =
1.454\times10^{19}$ cm for the 40kK benchmark model, as opposed to
1.463$\times10^{19}$ cm for the average radial distance). In real H~{\sc ii}
regions, dynamical effects dominate in these outer regions, where
ionized gas is flowing away from the ionization front. With 65 cells
on a side and the star at the centre, each cell has a width of 3\% of
the radius. Thus, the region we disregard (with hydrogen being
$\ge$25\% neutral) is much less than the width of each cell. 

We want to identify ``extreme" predictions, defined by being outside of
the range shown by the other models. One way of describing them
\citep{p01} is the ``isolation factor", the ratio of our value to the
next most extreme.

For the 40kK benchmark, our code provides three extreme values, two
within 4\% of the nearest value. Our [S\,{\sc III}] $\lambda$9532+9069
prediction is 12.5\% larger than that from the Rubin code (RR in the
table). Rubin's 
code (RR), {\sc mocassin} (BE), and Harrington's (PH) 
are the only ones that use detailed radiative transfer rather than an 
approximation such as ``outward only." 

For the 20kK benchmark, the results are very similar. We are extreme
in six quantities, by $<$5\% in all but two. The worst are [O{\sc
~III}] $\lambda$5949,5007 and (52$\mu$m + 88$\mu$m), (10\% and 8\%
more than T. Kalman's {\sc xstar}). We did not account for optical thickness
in these lines, so our high predictions are understandable. Also, for
this cool model our photon statistics for O$^{+2}$ are problematic, as
discussed above.  Since 1D codes can provide excellent spatial
resolution by inserting zones wherever there are steep gradients,
especially at the outer edges where the temperature is changing
rapidly, we feel that the benchmark results are satisfactory.

One simple check is available for the cool model. There are relatively
few He-ionizing stellar photons, and it is simple to show that He
absorptions strongly dominate H for He-ionizing photons. Each He
ionization is followed by the emission of 0.96 H-ionizing photons
\citep{o89}. Thus, the ratio of $\langle {\rm He}^+\rangle/\langle {\rm
H}^+\rangle$ is the ratio of the He-ionizing to H-ionizing stellar
luminosities, 0.49 for a $T$ = 20 kK blackbody. Our code predicts 0.47,
suggesting that our relatively coarse cell size is satisfactory for
predicting He$^+$/H$^+$ and other ionic abundances.

\begin{figure*}
\centerline{\psfig{figure=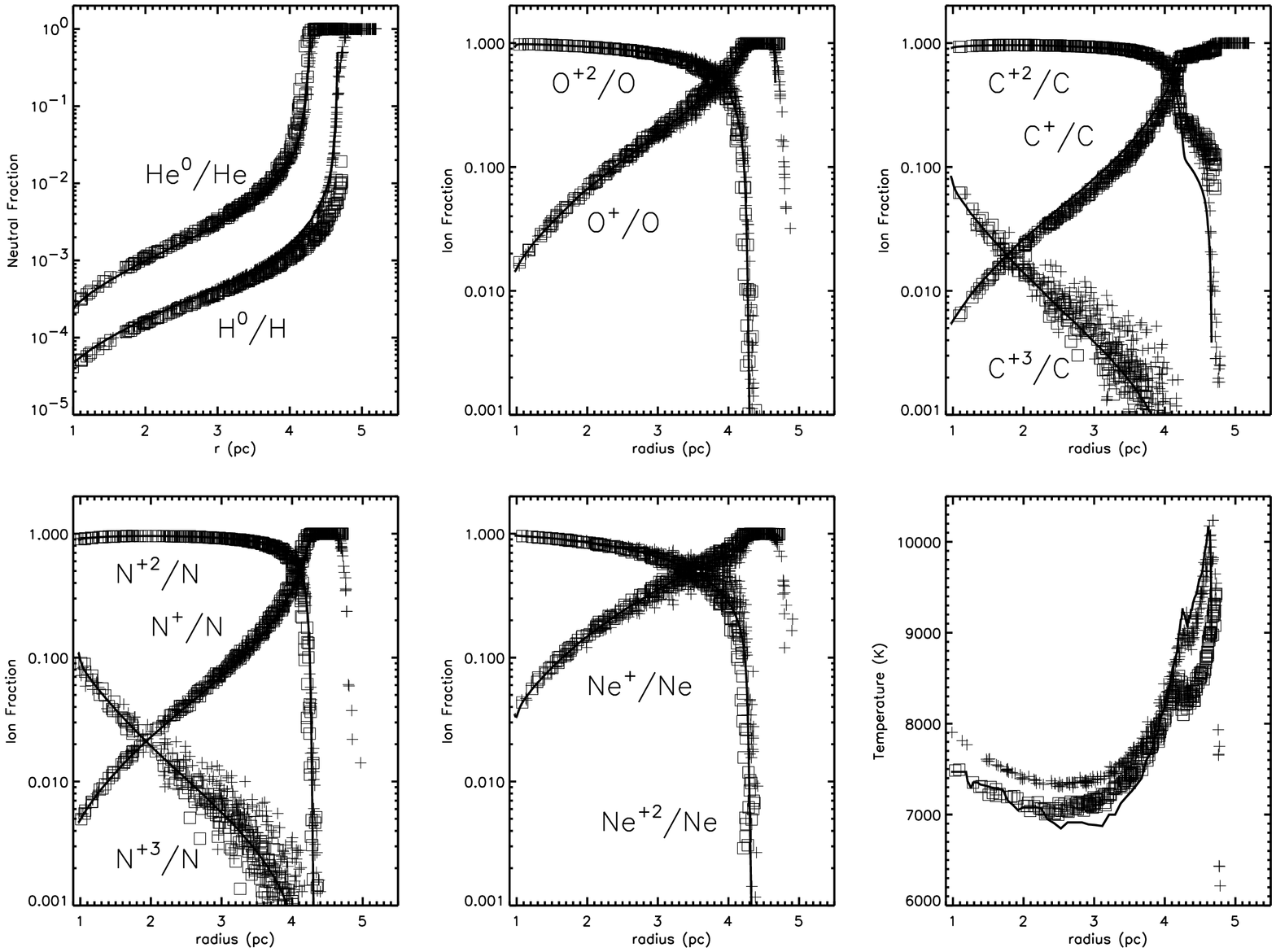,width=7.0truein,angle=90,height=5.0truein}}
\caption[]{HII40 benchmark, showing our Monte Carlo results (crosses), 
{\sc mocassin} (squares), 
and the {\sc cloudy} model (solid line).}
\end{figure*}

\begin{figure*}
\centerline{\psfig{figure=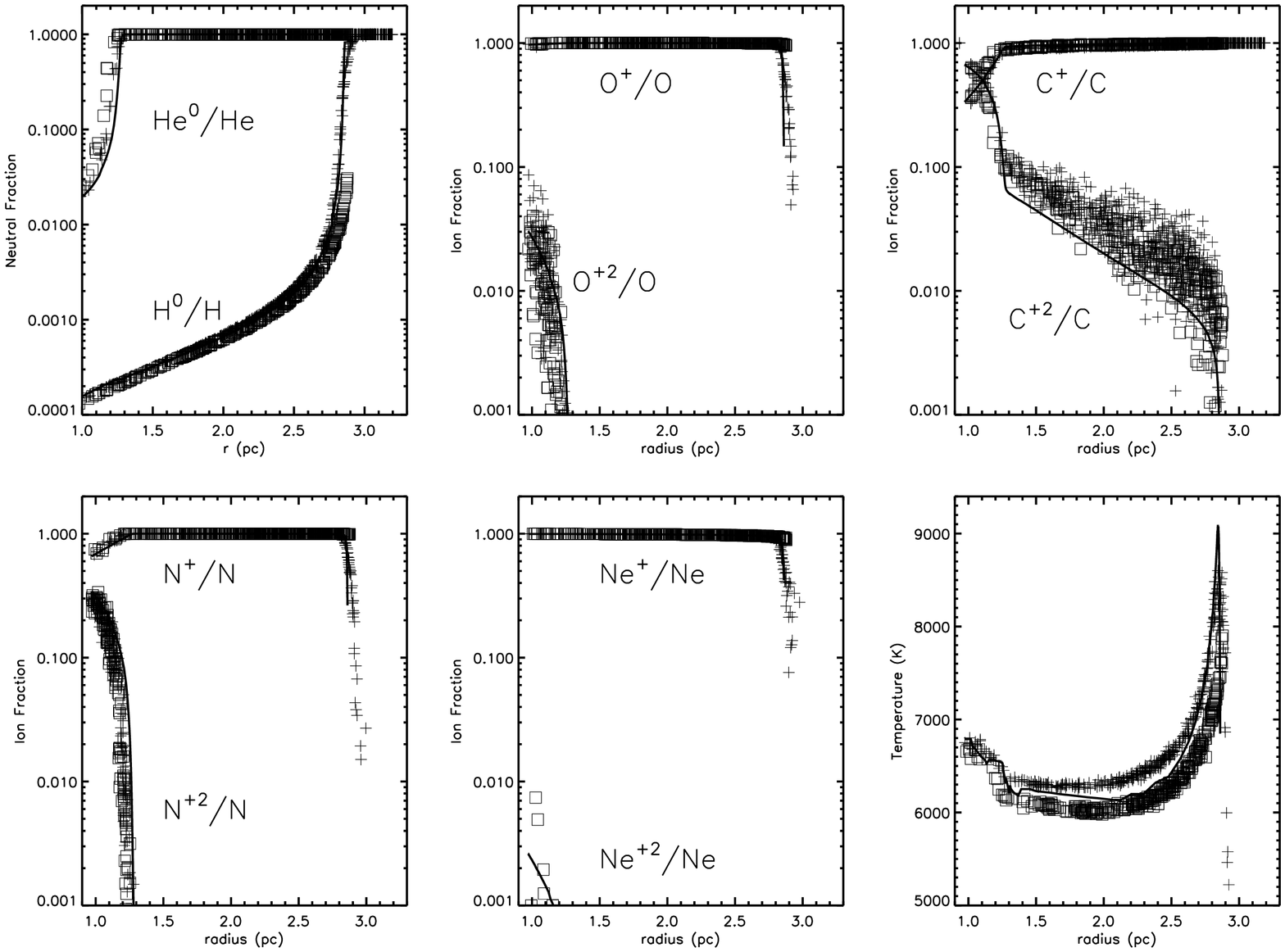,angle=90,width=7.0truein,height=5.0truein}}
\caption[]{HII20 benchmark, showing our Monte Carlo results (crosses), 
{\sc mocassin} (squares), 
and the {\sc cloudy} model (solid line).}
\end{figure*}

\begin{table*}
\begin{minipage}{115cm}
 \caption{Lexington H~{\sc ii} Region Benchmark Results: HII40$^a$}
 \label{symbols}
 \begin{tabular}{@{}lccccccccc}

 \hline
 Line/H$\beta$ & GF & HN & DP & TK & PH & RS & RR & BE& WME\\
 \hline
 ${\rm H}\beta/10^{36}{\rm erg\,s}^{-1}$
&2.06 &2.02 &2.02 &2.10 &2.05 &2.07 &2.05& 2.02 &2.01 \\
 He {\sc i} 5876
&0.119 & 0.112& 0.113& 0.116& 0.118& 0.116& -& 0.114 &0.114 \\
 C {\sc ii} 2325+
& 0.157& 0.141& 0.139& 0.110& 0.166& 0.096& 0.178& 0.148 &0.172 \\
C {\sc iii} 1907+1909
& 0.071& 0.076& 0.069& 0.091& 0.060& 0.066& 0.074& 0.041 &0.078 \\
 N {\sc ii} 122$\mu$m
& 0.027& 0.037& 0.034& -& 0.032& 0.035 & 0.030& 0.036& 0.031 \\
 N {\sc ii} 6584+6548
&0.669 & 0.817& 0.725& 0.69& 0.736& 0.723& 0.807& 0.852 &0.742 \\
 N {\sc ii} 5755
&0.0050 & 0.0054& 0.0050& -& 0.0064& 0.0050& 0.0068& 0.0061 &0.0057\\
 N {\sc iii} 57.3$\mu$m
&0.306 & 0.261& 0.311& -& 0.292& 0.273& 0.301& 0.223 &0.308 \\
 O {\sc i} 6300+6363
&0.0094 & 0.0086& 0.0088& 0.012& 0.0059& 0.0070& -& 0.0065 &0.011 \\
 O {\sc ii} 7320+7330
&0.029 & 0.030& 0.031& 0.023& 0.032& 0.024& 0.036& 0.025 &0.033 \\
 O {\sc ii} 3726+3729
&1.94 & 2.17& 2.12& 1.6& 2.19& 1.88& 2.26& 1.92 &2.23 \\
 O {\sc iii} 52+88$\mu$m
& 2.35& 2.10& 2.26& 2.17& 2.34& 2.29& 2.34& 2.28 &2.42 \\
O {\sc iii} 5007+4959
&2.21 & 2.38& 2.20& 3.27& 1.93& 2.17& 2.08& 1.64 &2.34 \\
 O {\sc iii} 4363
&0.00235 & 0.0046& 0.0041& 0.0070& 0.0032& 0.0040& 0.0035& 0.0022
&0.0044\\
Ne {\sc ii} 12.8$\mu$m
&0.177 & 0.195& 0.192& - & 0.181& 0.217& 0.196& 0.212 &0.192 \\
 Ne {\sc iii} 15.5$\mu$m
&0.294 & 0.264& 0.270& 0.35& 0.429& 0.350& 0.417& 0.267 &0.263 \\
 Ne {\sc iii} 3869+3968
&0.084 & 0.087& 0.071& 0.092& 0.087& 0.083& 0.086& 0.053 &0.063 \\
 S {\sc ii} 6716+6731
&0.137 & 0.166& 0.153& 0.315& 0.155& 0.133& 0.130& 0.141 &0.18 \\
 S {\sc ii} 4068+4076
&0.0093 & 0.0090& 0.0100& 0.026& 0.0070& 0.005& 0.0060& 0.0060 &0.0094 \\
 S {\sc iii} 18.7$\mu$m
&0.627 & 0.750& 0.726& 0.535& 0.556& 0.567& 0.580& 0.574 &0.623 \\
S {\sc iii} 9532+9069
&1.13 & 1.19& 1.16& 1.25& 1.23& 1.25& 1.28& 1.21 &1.44 \\
$10^3\times\Delta({\rm BC}\, 3645)$/{\AA}
&4.88 & -& 4.95& -& 5.00& 5.70& -& 5.47 &4.96 \\
 $T_{\rm inner}$/K
&7719 & 7668& 7663& 8318& 7440& 7644& 7399& 7370 &7743 \\
 $<T[{\rm N}_p{\rm N}_e]>$/K
&7940 & 7936& 8082& 8199& 8030& 8022& 8060& 7720 &8183 \\
 $R_{\rm out}/10^{19}$cm
&1.46 & 1.46& 1.46& 1.45& 1.46& 1.47& 1.46& 1.46 &1.45 \\
 $<{\rm He}^+{\rm H}^+>$
&0.787 & 0.727& 0.754& 0.77& 0.764& 0.804& 0.829& 0.715 &0.771 \\
 \hline

\multicolumn{10}{l}{$^a$GF, Ferland's {\sc cloudy}; 
HN, H. Netzer's {\sc ion}; DP, D. P\'equinot's {\sc nebu};
TK, T. Kallman's {\sc xstar};}\\
\multicolumn{10}{l}{PH, J. P. Harrington's code; RS, R. Sutherland's
{\sc mappings}; RR, R. Rubin's {\sc nebula}; }\\
\multicolumn{10}{l}{BE, B. Ercolano's {\sc mocassin}; WME, this paper.}
 \end{tabular}
\end{minipage}
\end{table*}

\begin{table*}
\begin{minipage}{115cm}
 \caption{Lexington H~{\sc ii} Region Benchmark Results: HII20}
 \begin{tabular}{@{}lccccccccc}
 \hline
 Line/H$\beta$ & GF & HN & DP & TK & PH & RS & RR & BE& WME\\
 \hline
 ${\rm H}\beta/10^{36}{\rm erg\,s}^{-1}$
&4.85 &4.85 &4.83 &4.90 &4.93 &5.04 &4.89& 4.97 &4.87 \\
 He {\sc i} 5876
&0.0072 & 0.008& 0.0073& 0.008& 0.0074& 0.0110& -& 0.0069 &0.00684 \\
 C {\sc ii} 2325+
&0.054 & 0.047& 0.046& 0.040& 0.060& 0.038& 0.063& 0.042 &0.053\\
 N {\sc ii} 122$\mu$m
&0.068 & -& 0.072& 0.007& 0.072& 0.071& 0.071& 0.071 &0.067\\
 N {\sc ii} 6584+6548
&0.745 & 0.786& 0.785& 0.925& 0.843& 0.803& 0.915& 0.846 &0.778 \\
 N {\sc ii} 5755
&0.0028 & 0.0024& 0.0023& 0.0029& 0.0033& 0.0030& 0.0033& 0.0025 &0.0025
\\
 N {\sc iii} 57.3$\mu$m
&0.0040 & 0.0030& 0.0032& 0.0047& 0.0031& 0.0020& 0.0022& 0.0019 &0.0049
\\
 O {\sc i} 6300+6363
&0.0080 & 0.0060& 0.0063& 0.0059& 0.0047& 0.0050& -& 0.0088 &0.055 \\
 O {\sc ii} 7320+7330
&0.0087 & 0.0085& 0.0089& 0.0037& 0.0103& 0.0080& 0.0100& 0.0064 &0.0089
\\
 O {\sc ii} 3726+3729
&1.01 & 1.13& 1.10& 1.04& 1.22& 1.08& 1.17& 0.909 &1.12 \\
 O {\sc iii} 52+88$\mu$m
&0.0030 & 0.0026& 0.0026& 0.0040& 0.0037& 0.0020& 0.0017& 0.0022 &0.0044
\\
O {\sc iii} 5007+4959
&0.0021 & 0.0016& 0.0015& 0.0024& 0.0014& 0.0010& 0.0010& 0.0011 &0.0026
\\
 Ne {\sc ii} 12.8$\mu$m
&0.264 & 0.267& 0.276& 0.27& 0.271& 0.286& 0.290& 0.295 &0.289 \\
 S {\sc ii} 6716+6731
&0.499 & 0.473& 0.459& 1.02& 0.555& 0.435& 0.492& 0.486 &0.521 \\
 S {\sc ii} 4068+4076
&0.022 & 0.017& 0.020& 0.052& 0.017& 0.012& 0.015& 0.013 &0.017 \\
 S {\sc iii} 18.7$\mu$m
&0.445 & 0.460& 0.441& 0.34& 0.365& 0.398& 0.374& 0.371 &0.381 \\
S {\sc iii} 9532+9069
&0.501 & 0.480& 0.465& 0.56& 0.549& 0.604& 0.551& 0.526 &0.602\\
 $10^3\times\Delta({\rm BC}\, 3645)$/{\AA}
&5.54 & -& 5.62& -& 5.57& 5.50& -& 6.18 &5.52 \\
 $T_{\rm inner}$/K
&7224 & 6815& 6789& 6607& 6742& 6900& 6708& 6562 &6852 \\
 $<T[{\rm N}_p{\rm N}_e]>$/K
&6680 & 6650& 6626& 6662& 6749& 6663& 6679& 6402 &6692 \\
 $R_{\rm out}/10^{18}$cm
&8.89 & 8.88& 8.88& 8.7& 8.95& 9.01& 8.92& 8.89 &8.73 \\

 $<{\rm He}^+{\rm H}^+>$
&0.048 & 0.051& 0.049& 0.048& 0.044& 0.077& 0.034& 0.041 &0.047 \\
 \hline
 \end{tabular}
\end{minipage}
\end{table*}

\subsection{Ionization in a Shadow Region}

The structure of the shadow zone behind a small high-density clump of
gas, such as those found in the planetary nebula NGC~7293, is of
direct astrophysical interest. There have been several treatments of
such a shadowed zone. Approximate static models have also been
considered by \citet{vba} and \citet{m76}. An analytical treatment is
given by \citet{c98}, who considered dynamical evolution of the
shadowed region and its ionization and temperature structure.  These
assume that the radiation field incident upon the shadow is the
diffuse field within the H~{\sc ii} region, estimated by assuming that
the nebula is thick to its own diffuse radiation {\em and that there
is no diffuse radiation produced by} He.

We tested the influence of He upon the ionization and temperature of a
shadowed zone by comparing two models, one with the benchmark
composition and one with the same H and heavy elements but no
helium. Both used the geometry similar to that in \citet{c98}: a flat
circular clump at $z$ = 0, completely opaque to incident
plane-parallel starlight out to its outer edge, so that it acts as a
circular plate blocking stellar radiation incident normally upon
it. We chose the uniform nebular density (100~cm$^{-3}$) so that the
stellar radiation is completely absorbed along the 6.2 pc $z$ axis of
the cube, even far from the clump. The ionizing spectrum is that of a
40000~K blackbody with an ionizing flux of $6.4\times
10^{10}$~cm$^{-2}$~s$^{-1}$. The radius of the occulting disk is 1.14
pc. We used periodic boundary conditions, meaning that when a photon
exits one ($x,y$) face of the cube it reappears on the other
side. Photons incident upon the $z$ = 0 face from within the cube are
reflected back into the cube. 

\begin{figure*}
\centerline{\psfig{figure=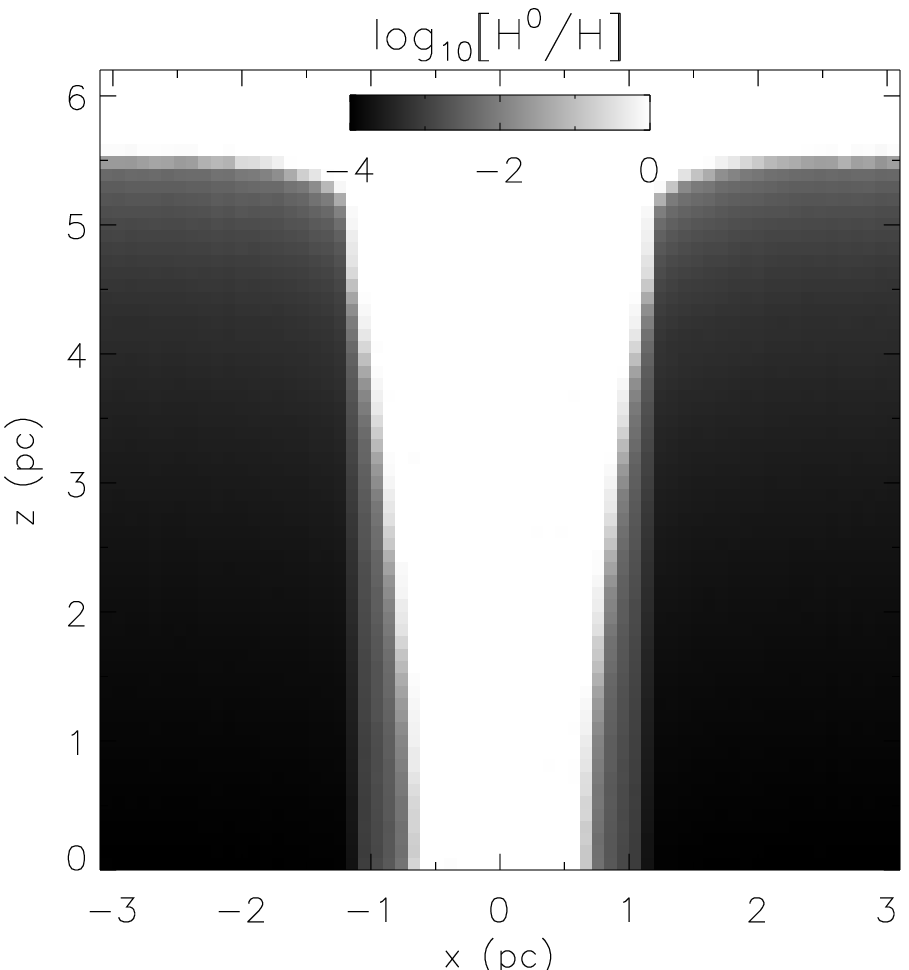}}
\caption{The ionization structure,  H$^0$/H, of a shadowed region
produced by an impervious circular cloud, with ionizing stellar
radiation incident from below as described in \S8.2. }
\label{tvshadow}
\end{figure*}

The ionization in the shadow is indicated by Fig. \ref{tvshadow} for
the standard benchmark composition. We see that the diffuse radiation
can penetrate relatively deeply into the shadow near the bottom of the
figure (near the plate); the incident intensity is large because H is
highly ionized. In this case, diffuse photons produced within
relatively large distances can reach the shadow. At large $z$, near
the top of the figure, the opacity in the directly illuminated
regions is relatively large, the nebular radiation into the shadow is
correspondingly low, and the ionized region has a sharp edge.

\begin{figure*}
\centerline{\psfig{figure=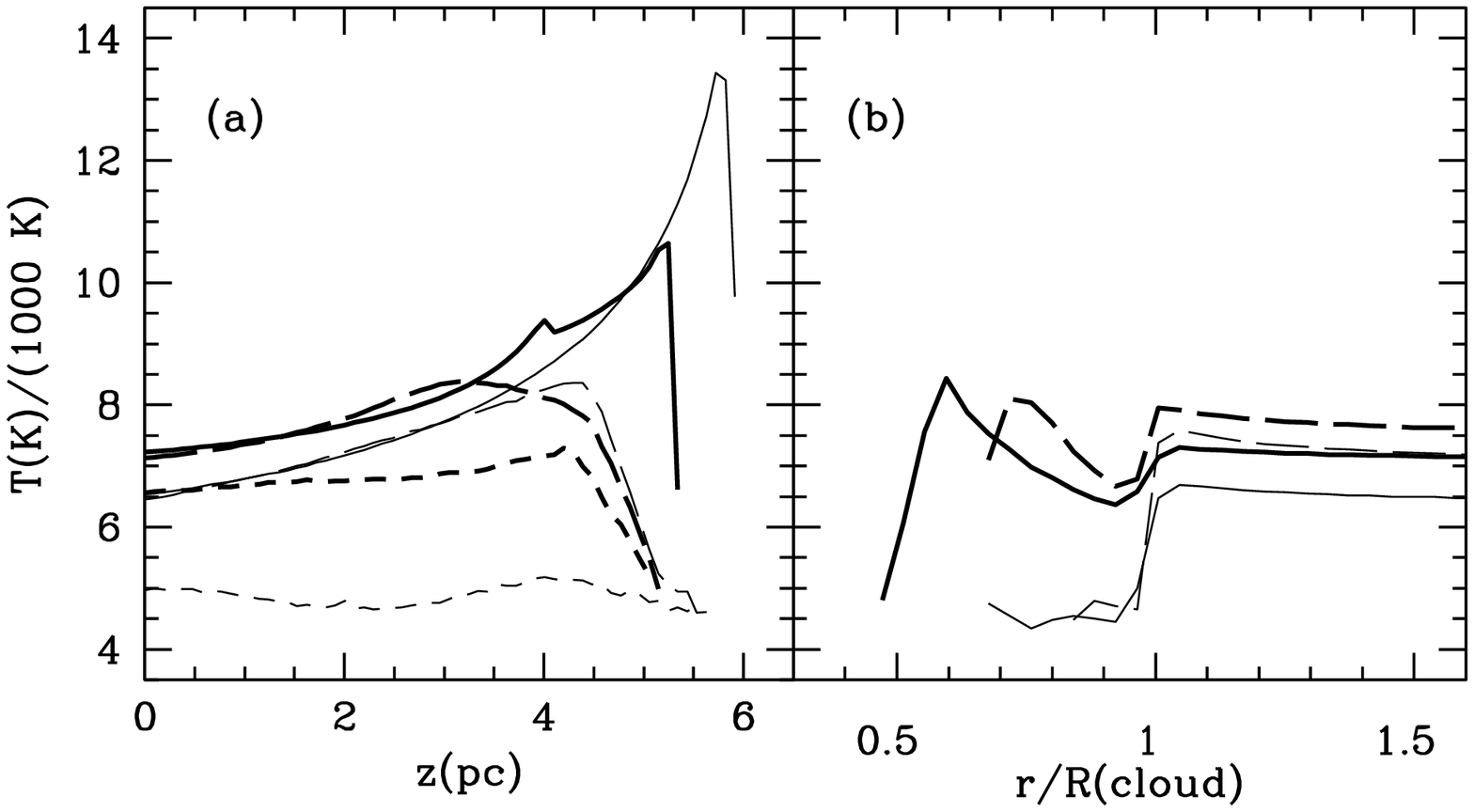}}
\vspace{-7.5cm}
\caption{(a) The nebular temperature, $T(r,z)$, within a region shadowed 
by a circular cloud that acts as an opaque plate, as shown in Fig. 3.
Stellar photons are injected parallel to the $z$ axis (vertical in
Fig. 3); $r$ is the radial distance from the axis of the shadow. The
radius of the cloud is 1.14 pc. Heavy lines show the case of the 
benchmark nebular composition, roughly solar; light lines are solar 
except that there is no helium.  
{\it Solid lines;} $r/R$(cloud) = 1.17; {\it long dashes:},
$r/R$(cloud) = 1.01; {\it short dashes:}, $r/R$(cloud) = 0.96. 
The figure shows the dramatic effects
that He recombination radiation has on the physical conditions within
the shadow (see text), most importantly, the very low $T_e$ if there
is no He. (b) $T(r,z)$ as a function of $z$ at $r/R$(cloud) = 0.095 pc 
(solid lines) and $z$ = 2.6 pc (dashed).  } \label{tvsrz}
\end{figure*}

Figure \ref{tvsrz} shows $T(r,z)$ vs. $r/R_{\rm cloud}$, the distance
from the axis of the shadow relative to the cloud's radius, and also
$T(r,z)$ vs. $z$, the distance along the shadow axis. Heavy lines are
for standard composition; light lines show the no-He models.

Fig. \ref{tvsrz}(a) shows $T(r,z)$, plotted against $z$, at $r/R_{\rm
cloud}$ = 1.17 (solid line), 1.01 (long dashes), and 0.964 (short
dashes). The lines terminate when H becomes $>$95\% neutral. We see
that for the inner no-He model (light short dashes), $T(r/R_{\rm
cloud} =0.964$) is much lower than the other temperatures at all $z$,
as predicted in \citet{c98}. H recombination radiation has a mean
energy of only $\sim0.68k_BT_e$ = 0.4 eV\footnote{In common notation,
the rate of kinetic energy loss per volume from recombinations is
$n_e\,n(\rm H^+)\,\beta_{\rm B}kT_e$, and the mean number of photons
produced is $n_e\,n(\rm H^+)\,\alpha_{\rm B}$. Values of these
functions are given in \citet{o89}.}. Photons released during the
cascade following He recombinations (mostly the 19.8 eV 2$^3\rm
S\rightarrow1^1$S line) can release $\sim(20-13.6)$~eV~=~6.3 eV in
photoelectric heating with each ionization and penetrate $\sim$3 times
as far as H recombinations in the shadow. Thus, the He recombination
photons have a major effect on the heating even though there are 10
times as many from H. Even just outside of the shadow (the long dashed
curves) the shadow temperature is about the same as far from the cloud
(the solid curve), until the He becomes neutral in the gas fully
illuminated by the star, at the point of the slight blip at $z\sim$3.9
pc in the heavy solid curve. Outside of the shadow, we see the
expected increase of $T(r,z)$ with $z$, showing the effect of
hardening the stellar radiation, for both helium- and no-helium
models. (Since we did not account for cooling by collisional
excitation of H lines, the maximum temperature reached by the no-He
model is probably significantly too large.) Within the shadow (short
dashes), there is little change of $T$ with $z$ because the spectrum
of the ionizing recombination radiation changes little along the axis
of the shadow. Within the shadow, $T(z)$ for the He models (heavy
short dashes) is significantly less than outside, so the collisionally
excited line strengths are appreciably weaker.

Fig. \ref{tvsrz}(b) shows $T(r,z)$ plotted against $(r/R_{\rm cloud}$)
at two values of $z$: solid lines are for $z$ = 0.1~pc; dashed for 2.6~pc. 
As before, the thick lines are for models with He, the thin for no
helium. The inward drop of $T(r)$ at $r=R_{\rm cloud}$ is obvious. We see that
the low $T_e$ of the no-He models holds at all $r$. There is an 
increase in $T(r)$ near the center for the helium models because the
outer part of the shadow is mainly ionized by the soft recombination
photons from H, while the He photons penetrate and deposit more energy
when they are absorbed. There is no similar effect for the no-He
shadow; all ionizing photons are soft. We see the penetrating power of
the He photons into the shadow. Outside the shadow
$T(r)$ decreases somewhat with increasing $r$ because the harder
diffuse photons can reach the shadow more easily than the softer.

This example shows the importance of diffuse helium radiation in
shadow regions and may also help explain why He is less ionized than H
in the Diffuse Ionized Gas \citep{rt}. It shows that any treatment of
the nebular diffuse radiation field inpinging onto shadowed regions
must include a careful treatment of the helium recombination
radiation. We will investigate this effect and other 3D
photoionization models in future papers.

An anonymous referee provided helpful comments. JSM thanks a PPARC
Visitors Grant to the University of St. Andrews, KW acknowledges support 
from a PPARC Advanced Fellowship.  We thank Kirk Korista
for comments and providing the {\sc cloudy} benchmark models and Mike Barlow
and Barbara Whitney for comments on early versions of the paper.
The {\sc mocassin} tests were carried out on the Enigma SunFire Cluster, 
at the HiPerSPACE Computing Centre, UCL, which is funded by PPARC.

\label{lastpage}

\end{document}